\documentclass[a4paper,10pt]{article}
\usepackage{inputenc}

\usepackage{hyperref}
\usepackage{amsmath}
\usepackage{graphicx}
\usepackage{amsthm, amssymb}
\usepackage{verbatim}
\usepackage{authblk}
\usepackage{longtable}

\usepackage{xcolor}

\newcommand{\Sbrackets}{S}
\newcommand{\gX}{\ensuremath{g}}
\newcommand{\gD}{\ensuremath{\widetilde g}}
\newcommand{\FuzzyR}{{\cal R}}

\title{Critical Behaviour of the Fuzzy sphere}
\author[1]{Denjoe O'Connor}
\author[1,2]{Brian P. Dolan}
\author[1]{Martin Vachovski}
\affil[1]{\hbox{School of Theoretical Physics}\\
\hbox{Dublin Institute For Advanced Studies}\\
\hbox{10 Burlington Rd, Dublin 4, Ireland;}}
\affil[2]{\hbox{Department of Mathematical Physics}\\
\hbox{National University of Ireland Maynooth}\\
\hbox{Maynooth, Ireland}}
\begin{document}

\maketitle

\begin{abstract}
We study a multi-matrix model whose low temperature 
phase is a fuzzy sphere that undergoes an evaporation transition
as the temperature is increased. 
We investigate finite size scaling of the system as the limiting 
temperature of stability of the fuzzy sphere phase is approached.
We find on theoretical grounds that the system
should obey scaling with specific heat exponent $\alpha=\frac{1}{2}$,
shift exponent $\overline{\lambda}=\frac{4}{3}$ and that the peak in
the specific heat grows with exponent $\overline{\omega}=\frac{2}{3}$.
Using hybrid Monte Carlo simulations we find good collapse of specific
heat data consistent with a scaling ansatz which give our best estimates 
for the scaling exponents as 
$\alpha=0.50\pm0.01$, $\overline{\lambda}=1.41\pm0.08$ and
$\overline{\omega}=0.66\pm0.08$. 
\end{abstract}

\section{Introduction}
\label{sec:Intro}

Different approaches to the basic structure of
spacetime exist and in recent years there has been a growing interest
in the notion of space or spacetime as an emergent concept.  A natural
setting where space or spacetime are necessarily emergent is
in the proposed nonpterturbative definitions of string theory provided
by matrix models \cite{BFSS,IKKT,BMN}. In this context plausible
models of emergent geometry have been discussed
\cite{Steinacker:2010rh}. The notion of classical geometry changes
drastically within the context of matrix models; neither background 
geometry nor topology is predefined but instead they emerge
dynamically as a consequence of the condensation of the matrix degrees
of freedom to form the background geometry.

The purpose of this paper is to study the scaling of finite matrix
effects for large matrix size (large $N$) as the limit of
stability of the geometry is approached.

We consider an action in which the basic objects are simple
Hermitian matrices at finite temperature with a prescribed energy
functional. The geometry exists at low temperatures as a condensate 
around which the system fluctuates.  

The energy functional of interest here is that of a 3-matrix model 
consisting of the trace of the square of the commutator of the 
matrices (a Yang-Mills term) plus the epsilon-tensor contracted with 
the trace of the three matrices 
(a Myers term \cite{myers}).  The system has been studied before in 
\cite{Azuma:2004zq,CastroVillarreal:2004vh,Azuma:2004ie,OConnor:2006,DelgadilloBlando:2007,DelgadilloBlando:2008vi} and with a mass deformation in \cite{DelgadilloBlando:2012xg}. It 
is a static bosonic subsector of the BMN model \cite{BMN}.

The model exhibits a geometrical phase for sufficiently low
temperatures where the effect of the cubic Myers term is important,
with the geometry being that of a fuzzy sphere
\cite{Madore:1991bw,hoppe} which is a non-commutative
version of the commutative sphere.

At a critical temperature, which can be traded for a critical Myers
coupling, a phase transition occurs and the condensed geometry
evaporates. In the geometrical, low temperature phase, small
fluctuations around this condensate correspond to a $U(1)$ gauge
and scalar field multiplet \cite{CastroVillarreal:2004vh}.

In previous studies
\cite{DelgadilloBlando:2007,DelgadilloBlando:2008vi} it was argued
that, as the transition is approached from the low temperature side,
there are divergent fluctuations in the system and in particular that
the specific heat diverges with exponent $\alpha=\frac{1}{2}$.  This
suggests that the system may exhibit finite size scaling
\cite{BarberinDombLebowitz:1983,Cardy_FiniteSizeScaling} in terms of
the matrix size as the system size grows.   We therefore study the growth of
fluctuations, in terms of both temperature and matrix size, as the
transition is approached from below, but insisting on remaining in the fuzzy
sphere phase.  The fluctuations we consider are those of a restricted ensemble
and do not take into account the rare finite matrix transitions when
the system jumps from the fuzzy sphere to the matrix phase. Such
fluctuations are completely absent in the large $N$ limit as is
typical of matrix models where tunneling goes to zero as $N$ goes to 
infinity. Our study shows that the fluctuations do indeed scale with 
matrix size.

The principal results of this paper are:
\begin{itemize}
\item In the absence of fluctuations, the Myer's term gives rise to an
  instability of the model. This is responsible for the
  destabilisation of the matrix phase for small matrices but
  insufficient to cause an instability for matrices of size $N\ge12$
  approximately.
\item The transition is rounded by finite matrix effects and there is
  a pseudo critical temperature shifted from the infinite $N$
  transition temperature with shift exponent which we predict on
  scaling grounds is $\overline{\lambda}=\frac{4}{3}$.
\item The scaling form of the free energy with $N$ is a universal
  function of $x=t N^{\overline\lambda}$ where
  $t=\frac{T_c-T}{T_c}$. We predict the scaling relation
  $\overline\lambda=\frac{2}{2-\alpha}$.
\item If we assume that $d=2$, based on a fuzzy sphere background, and
  $\alpha=\frac{1}{2}$ as found in earlier studies, then our theoretical
  prediction is that $\overline\lambda=\frac{4}{3}$ and if further we
  accept that the approach to criticality is governed by a divergent
  correlation length then we infer that the correlation length
  exponent $\nu=\frac{2}{3}$.
\item Hybrid Monte Carlo simulations give good agreement with finite size
  scaling in terms of the matrix size $N$ and assuming the asymptotic
  scaling form of the specific heat as the critical temperature is
  approached from below $C_v\simeq A_{-}(T_c-T)^{-\alpha}$ our best
  estimates from these measurements are $A_{-}=0.051\pm0.017$,
  $\alpha=0.50 \pm0.01$, $\overline\lambda=1.41\pm0.08$ and
  $\overline\omega=0.66\pm0.08$ where the peak in the specific heat
  grows as $A_{{\overline\omega}}N^{\overline\omega}$.  These are in
  good agreement with the theoretical estimates based on scaling and
  $\alpha=\frac{1}{2}$.
\item When we assume the exponent $\alpha=\frac{1}{2}$ and scaling so that $\overline{\lambda}=\frac{4}{3}$ and $\overline{\omega}=\frac{2}{3}$ we 
 find tight estimates for the shift amplitude $A_{\overline{\lambda}}=3.9\pm0.1$ 
 and the specific heat maximum amplitude $A_{\overline{\omega}}=0.199\pm0.005$.
\end{itemize}

The organization of the paper is as follows: In section \S 2 we discuss
the main properties of the model which are relevant to us. In section
\S 3 we present the theory of finite size scaling.  In section \S 4 we give
an overview of the difficulties that arise in near-critical system
simulations and the impact they have on our studies. In section \S 5 we
present our numerical results.  In section \S 6 we present our
conclusions from the study.

\section{The three matrix model}
\label{sec:TheModel}

The model we shall consider in this paper is the three matrix model,
which was studied in
\cite{Azuma:2004zq,OConnor:2006,DelgadilloBlando:2007,DelgadilloBlando:2008vi}.
Let $X_a$, $a=1,2,3$, be three traceless $N$-dimensional Hermitian
matrices.  We consider the action (really an energy divided by
temperature, as all our considerations will be in Euclidean signature)
\begin{equation}
 \mathcal{S}[X]=N Tr\Big(-\frac{1}{4}[X_a, X_b]^2 + \frac{2i\gX}{3}\epsilon_{abc}X_aX_bX_c\Big)
\label{eq:3MM_action}
\end{equation}
 where $\epsilon_{abc} $ is the totally antisymmetric Levi-Civita symbol,
$\gX \in \mathbb{R}$ is a parameter of the model.
The change $\gX \to -\gX$ is equivalent to $X_a\rightarrow -X_a$,
therefore it will be sufficient to restrict our study to the case $\gX \ge 0$,
which we shall assume. This model has a phase transition \cite{Azuma:2004zq,OConnor:2006} and it is the vicinity of this transition that is of interest to us here.

The stationary points of the system follow from varying
$\mathcal{S}$: demanding that $\delta \mathcal{S} = 0$ results in
\begin{equation}
 [X_b,[X_a,X_b] - i\gX \epsilon_{abc} X_c] = 0,
\label{eq:3MM_EM}
\end{equation}
and every configuration of matrices $X_a$ that solves \eqref{eq:3MM_EM} is a 
(local) extremum or saddle point of \eqref{eq:3MM_action}.

For most purposes of this paper it will be convenient to scale out a
factor of $\sqrt{N}$ and work with the parametrization $\gD =
{\gX}{\sqrt{N}}$, as this gives a phase diagram that does not
dependent on $N$.  So we make the substitution $X_a\to \gD
\frac{D_a}{\sqrt{N}}$ in terms of which the action reads
\begin{equation}
 \mathcal{S}[D]=\frac{\gD^{4}}{N} Tr\Big(-\frac{1}{4}[D_a, D_b]^2 + \frac{2i}{3}\epsilon_{abc}D_aD_bD_c\Big).
\label{eq:3MM_action_temp}
\end{equation}
From this it is clear that observables are symmetric under
$\gD\rightarrow -\gD$, or equivalently under $\gX \rightarrow -\gX$.
For $\gD\neq 0$ we can interpret $T = {\gD}^{-4}$ as a temperature for
the system.\footnote{Generally we prefer to discuss the physical
  properties of the system in terms of the temperature $T$, though it
  may be more convenient in some situations to use either $\gD$ or
  $\gX$.  Note that, as there are no dimensionful quantities in the
  action, $T$ here is dimensionless.}

Many of the physical properties are characterised by the expectation
value of the action $<\mathcal{S}>$, the specific heat of the system 
and the distribution of the eigenvalues of the matrices $X_a$. 
Our study here will focus on the first two.  
The specific heat is defined as
\begin{equation}
C_v = \frac{<( \mathcal{S} -<\mathcal{S}>)^2>}{N^2} = \frac{<\mathcal{S}>}{N^2}-\gD^{4}\frac{d}{d\gD^{4}}\Big(\frac{<\mathcal{S}>}{N^2}\Big).
\label{eq:Cv_definition}
\end{equation}
For convenience we shall define $\Sbrackets:=<\mathcal{S}>$, this is the internal
energy divided by the temperature.

The model has at least two phases, which we call the commuting 
matrix phase and the
fuzzy sphere phase, and
the above three quantities behave quite differently in these two phases.
In a semi-classical approximation \cite{DelgadilloBlando:2007} the 
phase transition occurs at
\begin{equation}
\gD_c = \Big(\frac{8}{3}\Big)^{\frac{3}{4}}\quad 
\Leftrightarrow\quad T_c = \Big(\frac{3}{8} \Big)^3\approx 0.05273.
\label{eq:critial_point}
\end{equation}
This is remarkably close to our numerical result
of $0.0531\pm 0.0003$ obtained in \S \ref{sec:Tc}.

\subsection{The commuting matrix phase}
\label{Sec:CommutingMatrixPhase}
The high temperature phase of the model with $T > T_c$
is characterised by fluctuations around a ground state
in which the three matrices are mutually commuting. 
This ground state can be represented by $X_a$'s which are
linear combinations of
$h H_m h^{-1}$, $m = 1,\dots, N-1$, where $H_m$ are in the Cartan sub-algebra
of $su(N)$ and $h \in SU(N)$.  Any such linear combination
is a trivial solution to \eqref{eq:3MM_EM}, so the classical action vanishes
for these stationary configurations.

However, these solutions can be unstable 
if any of the eigenvalues get too close to
one another, as we now demonstrate.
Fluctuations around a classical solution can be expressed as
\begin{equation}
X_a=X_{0,a}+\delta X_a,
\end{equation}
with $X_{0,a}$ three mutually commuting, Hermitian matrices.
We are free to perform an $N\times N$ unitary transformation on $X_{0,a}$ to simultaneously diagonalise them,
\begin{equation}
\bigl(X_{0,a}\bigr)_{ij} = \lambda^a_i \delta_{ij},\qquad \hbox{(no sum over $i$)}.
\end{equation}
A little algebra reveals that, to quadratic order in $\delta X_a$ we have
\begin{equation}
-\frac 1 4 Tr \Bigl [X_a,X_b]^2= 
\frac 1 2 \left( \bigl(\Delta_{ij}.\Delta_{ij}\bigr) \delta^{ab}
-\Delta^a_{ij}\Delta^b_{ij} \right) \delta X_{a,ij}\,\delta X_{b,ij}
\end{equation}
and
\begin{equation}
\frac {2i}{3} \epsilon_{abc}X_a X_b X_c = i \epsilon_{abc} \,\Delta^c_{ij}\delta X_{a,ij}\,\delta X_{b,ij},
\end{equation}
where $\Delta^a_{ij}=\lambda^a_i-\lambda^a_i$.
Stability of fluctuations around a classical solution are therefore determined
by the eigenvalues of the operator
\begin{equation}
\frac 1 2 \left( \Delta^2_{ij} \delta^{ab}
-\Delta^a_{ij}\Delta^b_{ij} \right) +i \gX \,\epsilon_{abc}\,\Delta^c_{ij}\,,
\label{MatrixFluctOP}
\end{equation}
where $\Delta_{ij}^2=\Delta^a_{ij}\Delta^a_{ij}$ with eigenvalues
$0$, $\frac 1 2 \Delta_{ij}^2\pm \gX \sqrt{\Delta^2_{ij}}$.  The
zero-eigenvalue is associated with the $U(N)$ invariance of the action
and can be removed by gauge fixing such that one of the matrices is
diagonal.  However, if some of the background eigenvalues, $\lambda^a_i$,
are too close together, (\ref{MatrixFluctOP}) has a negative eigenvalue, 
and hence an instability.
In particular, if
\begin{equation}\label{eq:instability_condition}
\Delta_{ij}^2 <  4\gX^2,
\end{equation}
for any pair $i$, $j$, then
there is a direction which is unstable.
The solution is stable if all the eigenvalues of $X_{0,a}$ are far enough apart.
Note that there are no unstable directions for $g=0$, the instability is
induced by the Myers' term.

Fluctuations can of course modify this analysis.
It is possible that they stabilise the unstable solutions. We will not 
attempt an analytic approach to this question here but will return to it 
later in the paper. The first immediate effect of fluctuations is
that they modify the expectation value of the action
and shift it away from $S = 0$.
To study this effect consider a Schwinger-Dyson type analysis,
\begin{eqnarray}
0&=& \int [DX] Tr \frac{\partial}{\partial X_a} \left(X_a e^{-\mathcal S}\right)\nonumber \\
\Rightarrow \qquad   0&=&3(N^2-1)- Tr < X_a  \frac{\partial \mathcal S}{\partial X_a}>\\
\Rightarrow\qquad 3(N^2-1)&=& - NTr<[X_a,X_b]^2> +2i\gX N \epsilon_{abc} Tr <X_aX_bX_c>\nonumber\\
&=& 4<\mathcal S> -2i\gX N \epsilon_{abc} Tr <X_aX_bX_c>,\nonumber
\end{eqnarray}
where $3(N^2-1)$ is the number of degrees of freedom in the three Hermitian 
matrices $X_a$.
Thus we expect
\begin{equation}
\frac{<{\cal S}>}{N^2} =\frac{3\bigl(N^2 -1\bigr)}{4 N^2} +\frac{i \gX}{2N} \epsilon_{abc} Tr <X_aX_bX_c>.
\end{equation}
It is shown numerically in \cite{DelgadilloBlando:2012xg} that 
$ Tr <X_aX_bX_c> \ \approx \frac {1}{N^{1/2}}\Bigl(\frac{1} {T^{1/4}} + 
o\bigl(\frac{1}{T^{1/2}}\bigr)\Bigr)$ at large $T$ and large $N$ 
so, in this limit, 
\begin{equation}
\frac{\Sbrackets_{m}(T)}{N^2} = \frac{3}{4} ,
\label{eq:action_exp_value_matrix}
\end{equation}
which is positive.
This suggests that, in the matrix phase of the model, the specific heat 
does not depend on $T$, 
\begin{equation}
  C_v = \frac{3}{4},
  \label{eq:Cv_value_matrix}
\end{equation}
and each degree of freedom contributes a value of $\frac 1 4 $ to the
specific heat.  The model behaves like a pure Yang-Mills matrix model
in the large $N$ limit, i.e. one with only the commutator squared
term.

For large $N$, the eigenvalue distribution in this phase is compatible
with a parabolic distribution \cite{Berenstein:2008eg,DelgadilloBlando:2012xg,O'Connor:2012vr}.  In a gauge in
which $X_3$ is diagonal (which can always be achieved by an $SU(N)$
transformation) the diagonal entries of $<X_3>$ can be arranged in
descending order and give a parabolic distribution with normalised
density
\begin{equation}
\rho(\lambda)=\frac {3(R^2 -\lambda^2)}{2\pi R^3},
\label{parabolicdistribution}
\end{equation}
with $R$ determined numerically to be $2.0$. This parabolic form of
the distribution further implies, as argued in
\cite{O'Connor:2012vr,Filev:2013pza}, that the background of commuting
matrices have their eigenvalues distributed uniformly within a ball of
radius $R=2.0$. Fluctuations around this background are still present and
those of the different matrices do not commute.

\subsection{The fuzzy sphere phase}
\label{FuzzySphere}

The fuzzy sphere phase is a cold (ordered) phase
and is radically different to the commuting
matrix phase.
The background matrices in this phase are represented by a 
solution to \eqref{eq:3MM_EM} in which $X_a$ are proportional
the generators of $su(2)$, $X_a = \gX L_a$
with $[L_a,L_b]=i\epsilon_{abc}L_c$,
up to $U(N)$ transformation $h L_a h^{-1}$ with $h \in SU(N)$. 
For the classical solution we have $\sum_a X_a^2 = \gX^2 c_2 \mathbf{1}$,
with $c_2$ the second order Casimir for the $N$-dimensional representation
of $SU(2)$,
and hence $\frac{<Tr D_a^2>}{N c_2}=1.$
More generally we shall define a radius of the fuzzy sphere, $\FuzzyR$, by
\begin{equation}
\FuzzyR^2 = \frac{<Tr D_a^2>}{N c_2},
\label{eq:FuzzyR}
\end{equation}
which, in the large $N$ limit, has a nonzero value only in the fuzzy
sphere phase. The parameter $\FuzzyR$ provides an order
parameter for the transition, being non-zero in the low temperature phase 
and zero in the high temperature phase.

For low temperatures below the transition the expectation value of the action 
in this phase is approximated by the value of the action for the 
solution $X_a = \gX L_a$, so that $\FuzzyR\simeq 1$ and 
\begin{equation}
\label{eq:action_exp_value_fuzzy}
\Sbrackets_{f}(T) = -\frac{c_2 c_2^{adj}}{12 T} + <\text{fluctuations}>
\end{equation}
where $c_2 = \frac{N^2 - 1}{4}$ and $c_2^{adj} = 2$ are the Casimir and adjoint Casimir operators of $su(2)$.
Since the matrices $X_a$ are proportional to the generators they have a discrete eigenvalue spectrum with
$N$ distinct eigenvalues of the form $\lambda = \Bigl\{-\gX\frac{N-1}{2}, -\gX\frac{N-3}{2}\ldots \gX\frac{N-1}{2}\Bigr\}$.

\subsection{Excited fuzzy sphere states}
\label{sec:ExcitedStates}
A closer examination of equation \eqref{eq:3MM_EM} shows 
that there are reducible fuzzy sphere solutions, with $X_a$ proportional to
$su(2)$ generators in a reducible representation of the form
\begin{equation}
R_1(M_1) \oplus R_2(M_2) \ldots \oplus R_K(M_k)
\label{eq:excited_states_form}
\end{equation}
where $R_i(M_i)$ is an $su(2)$ irreducible representation of dimension
$M_i$ and \hfill\break $\sum_{i=1, K} M_i = N$.  The matrices $X_a$
for this solution can always be chosen to have block-diagonal form and
this will be implicit for the rest of this discussion.  All of the
metastable states with the n-tuple $(M_1, M_2, \ldots, M_K)$ can be
listed and indeed all the solutions described so far --- even the
commuting matrix phase and the irreducible fuzzy sphere phase --- can
be classified this way.  
For a fixed $N$, the number of distinct solutions to (\ref{eq:3MM_EM}) of 
the form 
(\ref{eq:excited_states_form}) grows as $p(N)$, the number of integer 
partitions of $N$, which for large $N$ behaves as
\begin{equation}
p(N)\simeq\frac{e^{\pi\sqrt{2N/3}}}{4N\sqrt{3}}.
\label{eq:excited_states_count}
\end{equation}
and for low enough temperatures the fuzzy sphere configuration represents the ground state of the system with each of the other $p(N)-1$ configurations 
representing a potential metastable state. 

The special case $M_1 = M_2 = \ldots = M_N =
1$ corresponds to the commuting matrix phase and the ground state in
this case can be viewed as arising from $N$ one-dimensional (or
trivial) representations of $su(2)$. In the other extreme, when $K=1$,
the representation is irreducible and gives the fuzzy sphere discussed
above.  Nevertheless, the commuting matrix solution is genuinely
different to all the others, the configuration represented by diagonal
matrices is a state built from the one-dimensional representations so
the $\epsilon$ term in \eqref{eq:3MM_EM} plays no r\^ole and the
system has no memory of $su(2)$.  These are one dimensional
representations of an arbitrary algebra.

There are two observables that could be used to distinguish between the 
fuzzy sphere
from section \S\ref{FuzzySphere} and these excited configurations
with $K>1$.  First we can use the eigenvalues of the matrices: 
since any irreducible representation $R(M)$ 
of $su(2)$ has $M$ distinct eigenvalues, configurations of the form \eqref{eq:excited_states_form} have $max\{M_i\} < N$ distinct eigenvalues in their spectra.
Another observable that is sensitive to the excited states is the expectation value of the action. We have
\begin{equation}
\Sbrackets_{ef}(T, (M_1, M_2, \ldots, M_K)) = -\sum_{i = 1}^{K} \frac{M_ic_2(M_i) c_2^{adj}}{12 N T} + <\text{fluctuations}>
\label{eq:action_exp_value_excited_fuzzy}
\end{equation}
and $\Sbrackets_{f}(T) < \Sbrackets_{ef}(T, K) < \Sbrackets_{m}(T)$
for $\forall K: 1 < K < N$.  These excited states are unstable,
see section \S\ref{sec:ExcitedStates} and far from the phase transition 
fluctuations around the lowest excited states are very much like 
fluctuations around the ground state (see Fig. 
\ref{pic:excited_states_Cv_extraction}) and these fluctuations are small 
relative to the spacing between such states. So at very low temperatures 
these excited states do not play an important r\^ole in the 
thermodynamics of the system.

The considerations so far have been in the absence of fluctuations and
apply, at low temperatures, sufficiently far from the critical point
that fluctuations can be neglected. Fluctuations are important in the
high temperature phase due to eigenvalue repulsion which lifts the
degeneracy of eigenvalues and stabilises the high temperature phase against
the Myers instability discussed in section \S\ref{Sec:CommutingMatrixPhase}. 
Otherwise at high temperatures fluctuations are not large.
In the low temperature phase eigenvalue repulsion 
also lifts the degeneracy associated with 
identical $su(2)$ blocks in the low temperature phase otherwise for 
all practical purposes the excited fuzzy states play no 
r\^ole for sufficiently low temperatures.

However, near the critical point as the transition is approached from
the low temperature side fluctuations grow and the specific heat
rises, see Fig. \ref{pic:Cv_broad_temp_range}.  In this regime the
excited fuzzy sphere states will of necessity play a more important
r\^ole. One can estimate when the first excited state will be
important by noting that it corresponds to $R(N-1)\oplus R(1)$ and
that the difference in action between this and the ground state grows
linearly with $N$, being $\frac{N}{8}$ for large $N$.  If the square
of this difference divided by $N$ is larger than the specific heat
then these excited states are unimportant, however, as the critical
point is reached the specific heat grows and eventually all excited
states are important. Far from the transition $C_v=1$, so we can
estimate that the first excited state will begin to become important
for $T\sim \frac{1}{64}\simeq 0.0156$.  Once the first excited state
becomes important, there are more possibilities for fluctuations in
the system and the specific heat in turn grows. More excited states
enter the picture and eventually the system undergoes a phase
transition. Earlier estimates
\cite{Azuma:2004zq,CastroVillarreal:2004vh,Azuma:2004ie,OConnor:2006,DelgadilloBlando:2007} 
give this transition at $T_c={(\frac{3}{8})}^3\simeq 0.0527$. So we expect the
critical regime between for $0.0156\leq T\leq 0.0527$ which is quite
consistent with Fig. \ref{pic:Cv_broad_temp_range}.

\subsection{A 1.5 order phase transition}
\label{sec:1.5odder}

It is clear from the previous sections that the commuting
matrix and the fuzzy sphere phases are quite different. 
The classical results in equations
\eqref{eq:action_exp_value_matrix} and \eqref{eq:action_exp_value_fuzzy}
give $\Sbrackets_{m}(T_c) - \Sbrackets_{f}(T_c) \ne 0$
at the transition temperature, $T_c$,
so this might na\"{\i}vely be classified as a first order phase transition ---
with latent heat and a finite specific heat on either side of the transition ---
but the full story is more subtle.

An approximate analytic expression for the specific heat, 
in the $N\rightarrow \infty$ limit, was given in \cite{DelgadilloBlando:2008vi}.
If we make the ansatz $X_a = \phi\, \gX \,L_a$ in the fuzzy sphere phase
and write an effective potential for the theory in terms of $\phi$
then, in a large $N$ semi-classical approximation, 
equations (3.25) and (3.26) of reference
\cite{DelgadilloBlando:2008vi} with $m=0$, give, $\FuzzyR=\phi$ and 
in the large $\gD$ limit, 
\begin{equation}
C_v=\frac 3 4 + \frac{\gD^5 \phi^2}{32}\frac {d\phi}{d\,\gD}
\qquad
\hbox{with}
\qquad
\phi=1-\frac {2}{\gD^4} - \frac {12} {\gD^8} + o\left(\frac 1 {\gD^{12}}\right).
\label{CeeVee}
\end{equation}
Thus
\begin{equation}
C_v=1+\frac {2} {\gD^4} + o\left(\frac 1 {\gD^8}\right)
\end{equation}
and 
\begin{equation} C_v \quad
\mathop{\longrightarrow}_{T\rightarrow 0} \quad 1\,.
\label{eq:CvlowT}\end{equation}
On the other hand, near $\gD_c$
\begin{equation}
  C_{v}(\gD)= \left\{ 
      \begin{array}{l} \frac{29}{36}+\frac{1}{4\sqrt{6}}\sqrt{\frac{\gD_c}{\gD - \gD_{c}}}+\dots
\\
    \frac{3}{4}
    \end{array} \right. \phi=\left\{\begin{array}{ll}\frac{1}{4}+\sqrt{\frac{3}{8}}\sqrt{\frac{\gD-\gD_{c}}{\gD_{c}}}+\dots \,, &  \gD > \gD_{c}\,\\
0 \, &  \gD < \gD_{c}\, .
\end{array}\right.
  \label{eq:Cv_critical_behavior}
\end{equation}
Thus the specific heat diverges\footnote{In terms of temperature (\ref{eq:Cv_critical_behavior}) gives $C_v(T)=\frac{29}{36}+\frac{3}{64}(T_c-T)^{-\frac{1}{2}}$+\dots .} on the low temperature side of the transition. 
This is the characteristic behaviour of a continuous (also called 2nd order) 
transition near a critical point. 

The general theory of continuous phase transitions and critical phenomena
\cite{Pelissetto:2000ek,WidomB,Stanley:1999zz,Goldenfeld:1992qy}
suggests that, near the phase transition, the specific heat $C_v$ on
either side of the transition should behave as
\begin{equation}
 C_{v}(T) \sim C_{0\pm} + A_{\pm}  |T-T_{c}|^{-\alpha}\,.
\label{eq:CvT}
\end{equation}
The three matrix model under consideration here seems to have a rather
unusual phase transition in that the specific heat diverges only as 
the phase transition is approached
from one side but does not diverge as it is approached from the other.

\begin{figure}
\centering
\includegraphics[scale=0.95]{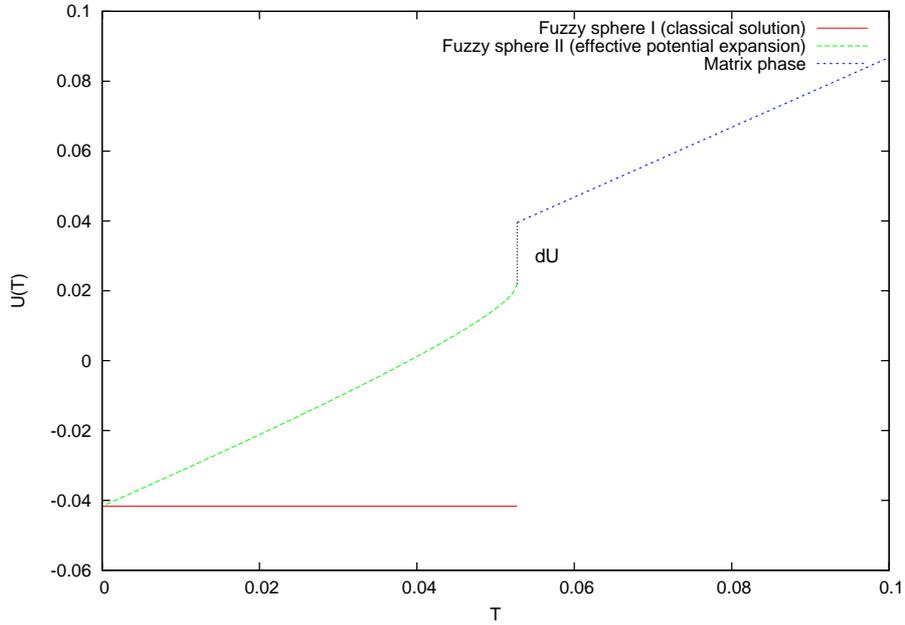}
\caption{The internal energy per degree of freedom showing critical
  behaviour on the low temperature side and non-critical behaviour on
  the high temperature side.  The slope near $T_c$, gives the specific
  heat.  }
\label{pic:internal_energy}
\end{figure}

The internal energy per degree of freedom, 
$U=T\frac{<\mathcal  S>}{N^2}$, arising from the semi-classical 
approximation of \cite{DelgadilloBlando:2008vi}, is plotted in
Fig.~\ref{pic:internal_energy} and the slope of this curve near $T_c$,
when expressed in terms of $\gD$, results in the form
\eqref{eq:Cv_critical_behavior} for the specific heat.  The
semi-classical approximation on the low-temperature side is given by
\begin{equation}\label{eq:semi-classical-S}
  \frac{S}{N^2} = \frac{3}{4} - \frac{\phi^3(T)}{24 T}
\end{equation}
where 
\begin{align}\label{eq:phiT}
  \phi(T) = \frac{1}{4}\Bigg( 1 + \sqrt{1 + \delta(T)} + \sqrt{2 - \delta(T) + \frac{2}{\sqrt{1 + \delta(T)}}} \Bigg) \\
   \delta(T) = 4T^{1/3}\Bigg(\Big(1 + \sqrt{1 - \frac{T}{T_c}}\big)^{\frac{1}{3}} + \big(1 - \sqrt{1 - \frac{T}{T_c}}\Big)^{\frac{1}{3}} \Bigg).\label{eq:deltaT}
\end{align}
This is the
typical behaviour of a critical point and a second order phase
transition.  This implies that a small correction to $T_c$ can give a
very large correction to the internal energy, $\Delta U_{T_c}$.  

On the high temperature side the internal energy is $U=\frac 3 4 T$,
from (\ref{eq:action_exp_value_matrix}), and so approaches the phase
transition with a finite slope, giving constant specific heat
(\ref{eq:Cv_value_matrix}).

This transition has the characteristic features of a 2nd order
transition when approached from low temperatures while those of a 1st
order transition when approached from the high temperature side.  We
might call such a transition a 1.5 order phase transition.  The two
dimensional dimer model has similar asymmetric thermodynamics in the
neigbhourhood of its transition. Curiously in the dimer example the
background geometry can also be interpreted as undergoing a transition
\cite{Dimers}.

\begin{figure}
\centering
\includegraphics[scale=0.8]{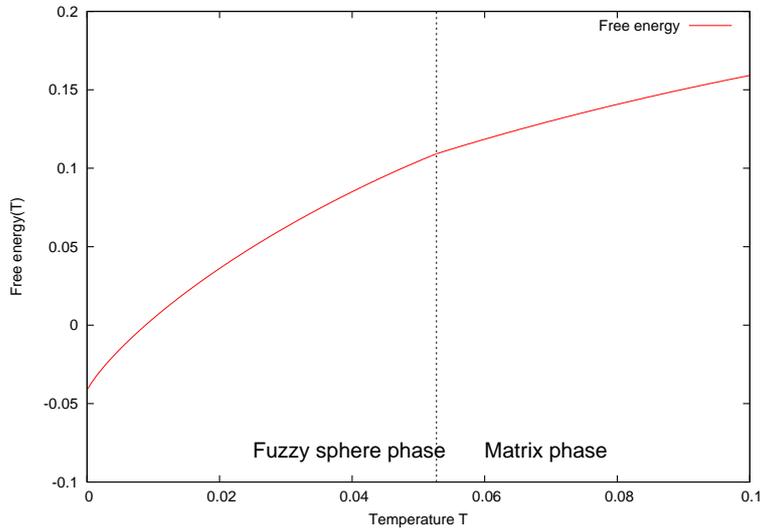}
\caption{A semi-classical approximation for the free energy per degree
  of freedom.  By construction the free energy is continuous at the
  critical point, but it is not differentiable, the left derivative is
  slightly greater than the right derivative, leading to the jump in
  entropy described in the text.}
\label{pic:free_energy}
\end{figure}

The free energy per degree of freedom was also derived, in the same approximation
as the internal energy above, in \cite{DelgadilloBlando:2008vi}.
On the low temperature side it is\footnote{
One must be careful in specifying the
measure when determining the free energy, and the measure for the matrices $X_a$ differs from that for the $D_a$ by a temperature dependent factor \cite{DelgadilloBlando:2008vi}.  The from of the free energy quoted here is that associated with the $D_a$.}
\begin{equation}
\frac{F}{N^2}=
T\bigg[\ln\left(\frac {\phi} {T} \right) - \frac 1 3\biggr]  -
\frac{\phi^4}{24}. \label{eq:Fformula1}
\end{equation}
Conversely, on the high temperature side, integrating 
\begin{equation}
U(T)=-T^2\frac{d}{d T}\left(\frac F T \right)=\frac 3 4 T,
\label{eq:UFrelation}
\end{equation}
leads to
\begin{equation}
\frac {F} {N^2}= C_1 T - \frac 3 4 T\ln T,\label{eq:Fformula2}
\end{equation}
with $C_1$ an integration constant.
Adjusting $C_1$ so that $F(T_c)$ matches on the high and
low side gives $C_1=\frac{\ln 6}{4} -\frac {7}{12}\approx -0.1354$ and results in the free energy per degree of freedom shown in
Fig.~\ref{pic:free_energy}.
There is a jump in the specific entropy\footnote{Note: There are $3(N^2-1)$ degrees of freedom.}, $s=-\frac 1 {3N^2}\frac {dF} {dT}$
as we go through the phase transition, $\Delta s = \frac {1} {9}$.

Of course this classical and semi-classical analysis is not the whole story
and indeed the purpose of the present work is to study the characteristics of this phase transition numerically.

\section{Finite size critical systems}
\label{sec:FiniteSize}

\begin{figure}
\centering
\includegraphics[scale=0.95]{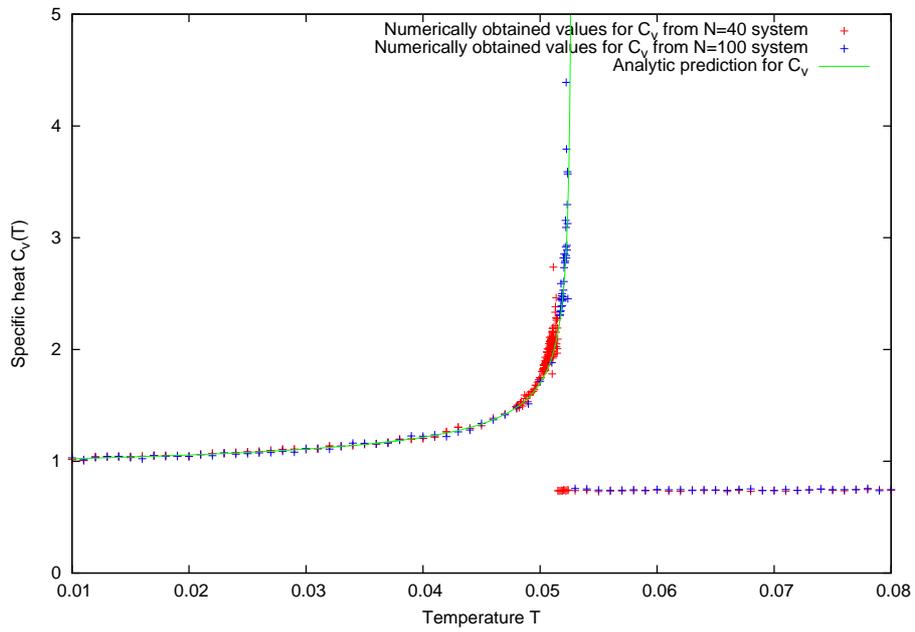}
\caption{Specific heat for $N=40$ and $N=100$ as a function of
  temperature (or equivalently, $\gD^{-4}$), compared to the
  theoretical prediction \eqref{CeeVee}.  The deviation of the
  numerical data from the theoretical curve, for largest values of
  $C_v$ plotted very near the critical point, is a finite size
  effect. We can also see that the maximum observable $C_v$ for
  $N=100$ is closer to the transition temperature than that for
  $N=40$. This is the shift in pseudo critical temperature.}
\label{pic:Cv_broad_temp_range}
\end{figure}

\begin{figure}
\centering
\includegraphics[scale=0.95]{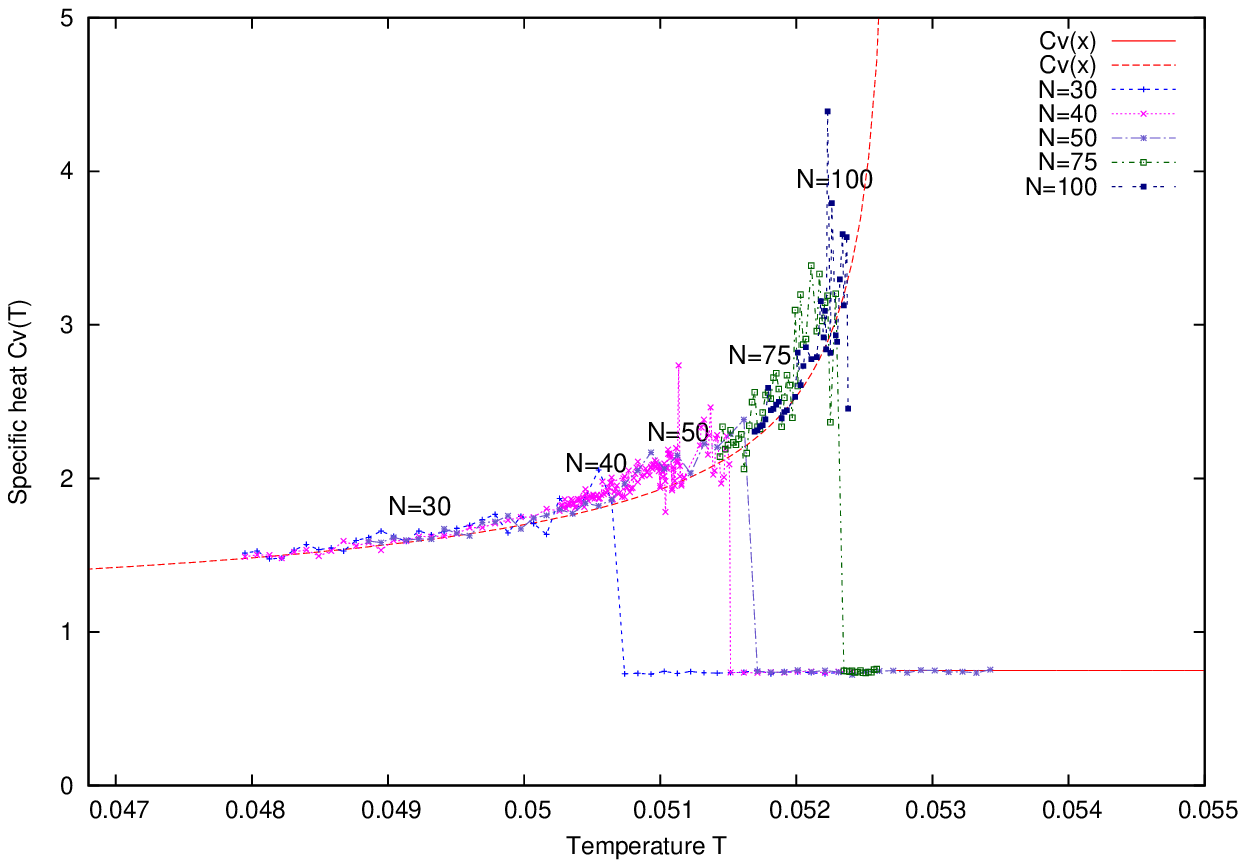}
\caption{Blow up of specific heat for different system sizes, compared to the theoretical prediction
\eqref{CeeVee}.}
\label{pic:Cv_scaling}
\end{figure}

Phase transitions where some observables are non-analytic functions of
the temperature, $T$ (e.g. they may diverge) are possible only in the
thermodynamic limit, which in our case would correspond to taking
$N\to\infty$ at fixed $T$.  We can of course only perform numerical
studies of systems consisting of a finite number of degrees of
freedom, so the systems we simulate will only undergo \textit{pseudo}
phase transitions where the non-analyticities are rounded (e.g.  with
peaks at pseudo-critical points instead of divergences).\footnote{For
  more a comprehensive treatment of critical phenomena the reader is
  directed to see the reviews in \cite{Pelissetto:2000ek, WidomB,
    Stanley:1999zz, Goldenfeld:1992qy}.}  Increasing $N$ gets closer
to the thermodynamic limit but also increases the computer resources
required for the numerical study, truly $N\to\infty$ systems can only
be approximated by finite $N$ systems and the thermodynamic limit must
be extrapolated from finite $N$ results.  The behaviour of the
specific heat as a function of temperature is plotted in
Fig.~\ref{pic:Cv_broad_temp_range}, for $N=40$ and $N=100$.  The
deviation, in the numerical data, of the largest values of $C_v$ for
different values of $N$ is the due to finite size effects. In
Fig.~\ref{pic:Cv_broad_temp_range} we see that the peak of the
specific heat moves with $N$, this is the shift of the pseudo-critical
temperature. The temperature at which the different curves begin to
deviate from one another corresponds to the rounding temperature and
is more difficult to observe in the figure.  In
Fig.~\ref{pic:Cv_scaling} we show a blow-up of the area around the
critical point for different values of $N$ where one sees the onset of
rounding more clearly. It is probable that the data shown in Figs.
\ref{pic:Cv_broad_temp_range} and \ref{pic:Cv_scaling} do not
achieve the true maximum specific heat since the data do not track
the return to low values of the specific heat at high
temperatures. However, this is not important for our analysis as we
will show that the entire critical regime satisfies scaling with the
matrix size.

Before going on to discuss our numerical results in detail we give a
brief review of finite size effects on critical systems. See 
C.~Domb and J.L.~Lebowitz \cite{BarberinDombLebowitz:1983}, for further 
discussion of these issues and for the original literature see
also \cite{Cardy_FiniteSizeScaling}.

\subsection{Thermodynamic limit away from the critical point}
\label{sec:ThermodynamicLimit}

Let $F(T, N_d)$ be the free energy for a system with $N_d$ degrees of 
freedom at a temperature $T$. In the thermodynamic limit the free energy 
per degree of freedom is 
\begin{equation}
  f_{\infty}(T) = \lim_{\hbox{$\scriptstyle N_d\to\infty$}} \frac{1}{N_d} F(T, N_d) .
\end{equation}
In our case $N_d = 3(N^2 - 1) \sim N^2$.  Far away from any
critical point we expect this limit to exist and to be independent of
the macroscopic geometry.
 
When the number of degrees of freedom is finite and all degrees of
freedom are equivalent, e.g. there are no surfaces and the couplings
are isotropic, then there is a characteristic linear system size
$L\sim a N_d^{\frac 1 d}$, where $d$ is the dimension and $a$ is a
microscopic scale such as a lattice spacing and the volume of the
system is $V=L^d$.  Since $a$ is fixed, and typically absorbed in the
parameters of the system, one can take $a=1$.

\subsection{Finite size effects and the correlation length}
\label{sec:CorrelationLength}

When a finite system approaches a critical regime there is a number
of important effects that must be taken into account.  First there is
a temperature, called the \textit{rounding temperature} and denoted
$T^*(L)$, at which observables of the finite system start to deviate
from those of the infinite system and $T^*(L) \to T_c$ as the system
size is increased.  Finite size scaling assumes that this approach of
the rounding temperature to the bulk critical temperature is governed
by scaling with
\begin{equation}
 |T^{*}(L) - T_c|/T_{c} \sim A_\theta  L^{-\theta}
  \label{eq:theta_definitionL}
\end{equation}
where $\theta$ is the rounding exponent.

Another finite size effect, which is directly visible in
Fig.~\ref{pic:Cv_scaling}, is that thermodynamic quantities which
diverge at the critical point merely have maximal values in 
finite systems with the maximum at some temperature $T_m(L) \ne T_c$.  Such
temperatures are called \textit{pseudo-critical}.  Again 
$T_m(L) \to T_c$, as the system size is increased 
and finite size scaling conjectures that 
\begin{equation}
  |T_{m}(L) - T_{c}|/T_{c} \sim A_{\lambda}  L^{-\lambda}
  \label{eq:lambda_definitionL}
\end{equation}
where $\lambda$ is the \textit{shift exponent}.

The specific heat of many critical systems is one such observable and 
finite size effects round a divergent specific heat so that it has a maximum,
\begin{equation}
C_{vm}(L) = C_{v}\big(T_{m}(L)\big) ,
\end{equation}
with $C_{vm}(L) \to \infty$ as $L\to\infty$.
Finite size scaling implies that the divergence emerges in the limit of  
infinite $L$  via scaling. Thus 
\begin{equation}
   C_{vm}(L) \equiv C_v(T_m(L)) \sim A_{\omega} L^{\omega}.
  \label{eq:omega_definitionL}
\end{equation}
The exponents 
$\theta$, $\lambda$ and $\omega$,
describe the critical behaviour of our finite system as the system size goes 
to infinity.  

As the critical point of a bulk system is approached the length scale,
over which fluctuations in the system are correlated, grows.  The
correlation length is defined as the rate of the asymptotically
exponential decay of the two-point function, with distance, $r$, between
the points. For the near-critical system
\begin{equation}
  \Gamma(r, T) \sim \exp(-r / \xi(T) )\quad \hbox{as} \quad r\to\infty\,,
\end{equation}
from which the $\xi(T)$ can be computed as
\begin{equation}
  \xi^{-1}(T) = -\lim_{r\to\infty} \frac{\ln\Gamma(r, T)}{r} .
  \label{eq:corr_length_limit}
\end{equation}
At a critical point the correlation length diverges and, in most
systems, in the immediate vicinity of the critical point $\xi(T)$ can
be expressed as\footnote{In fact there are systems where the
  correlation length diverges faster then any polynomial.  A famous
  example is the \textit{Kosterlitz-Thouless Phase Transition} where
  $\xi(T)$ diverges exponentially \cite{JMKosterlitz_DJThouless}}
\begin{equation}
  \xi(T) \sim f^{\pm} \vert t\vert^{-\nu}\qquad\hbox{where}\qquad
t=\frac{T-T_c}{T_c} 
  \label{eq:corr_length_exponent}
\end{equation}
is the reduced temperature and the $f^+$ and $f^-$ are system
dependent amplitudes above and below the critical temperature. They
are typically different from one another but their ratio is universal
within a given universality class.  The exponent $\nu$ therefore
dictates how fast the correlation length diverges when $t\to 0$ in an
infinite system.

The correlation length, of a system plays a crucial r\^ole
in the explanation of finite size scaling for systems exhibiting
critical phenomena such as the Ising model or gas systems.\footnote{In
  the case of matrix models, due to the non-local type of interaction
  between the entries and the absence of a notion of distance between
  the elements, we can only speculate on the existence of a unique
  correlation length.}
The finite size system correlation length $\xi_{L}$ is constrained by 
the size of the system and one expects that 
\begin{equation}
  \xi_{L}(T_{m}(L)) \sim L\,.
  \label{eq:rounding_temp_estimation}
\end{equation}
We can classify the different regimes for the system in terms of the scaled
variable $y=\frac{\xi_L}{L}$ or $z=\frac{\xi(t)}{L}$.  
As we approach criticality for finite size system the correlation length
takes the form 
\begin{equation}
\xi(T,L)=L y(z)=L f(t L^{1/\nu}). 
\end{equation}
where both $y$ and $f$ are universal scaling functions\footnote{The non 
universal scale for the argument of $f$ must be adjusted by convention to 
get all systems in a universality class to match. 
Such nonuniversal constants in scaling functions are referred to as  ``metric
factors'' and they depend on the microscopic details of the system and
are fixed by some system independent convention. Metric factors will not be 
important for our purposes here, since we have only one system.}.

In this critical regime and when the system is large relative to the
correlation length $y \ll 1$, $y(z)\simeq z$ and 
$f(x)\sim x^{-\nu}$ so that bulk scaling is recovered.  As the critical
temperature is approached finite size effects become important and
$y(z)$ begins to deviate from $z$.  This occurs at the rounding
temperature $T^*(L)$. For a system of finite extent the correlation
length cannot grow arbitrarily large relative to the system size and
at the pseudo-critical temperature the specific
heat reaches its maximum value. We take the maximum of the specific
heat to define the pseudo-critical temperature and at this temperature
we must have that both $y$ and $f$ are $L$ independent constants. 
Therefore $t_m L^{1/\nu}= const$, which implies
\begin{equation}
\vert T_m(L)-T_c\vert\sim L^{-{1/\nu}} .
\end{equation}
so that the prediction of finite size scaling is therefore that the 
shift exponent $\lambda=\nu^{-1}$.

If we assume that the only relevant quantity in an expansion around
$T_c$ is the correlation length, then finite size scaling implies that
the free energy, in the vicinity of the critical point, takes the form
\begin{equation}
\lim_{N\rightarrow\infty,T\rightarrow T_c}F(T,N_d)-N_df_{\infty}(T_c)
={\cal F}(t L^{1/\nu}) ,
\label{ScalingFreeEnergy}
\end{equation}
where ${\cal F}(x)$ is a universal scaling function that depends only on 
the universality class of the system. 
Given this form of the free energy we expect that the exponents 
$\lambda$ and $\theta$ should be the same, though the amplitudes
$A_\theta$  and $A_\lambda$ may differ.

If we take the large $L$ limit for fixed $t$, from the extensivity of the 
system we must get 
\begin{equation}
{\cal F}(t L^{1/\nu})\sim L^d t^{2-\alpha} 
\label{ScalingcalF}
\end{equation}
which requires ${\cal F}(x)\sim \vert x\vert^{2-\alpha}$, and we infer that
$2-\alpha=\nu d$ which is a well known scaling exponent relation.  

We can further take two derivatives of (\ref{ScalingcalF}) with
respect to $t$, and divide by $N_d$, to obtain the specific heat in
the scaling regime and using $2-\nu d=\alpha$ we obtain
\begin{equation}
C(T,L)=A_{-}L^{\alpha/\nu} {\cal C}(x)
\end{equation}
where the amplitude $A_-$ is extracted to guarantee that ${\cal C}$ is a 
universal scaling function. With $x$ the $L$ independent 
constant $x_m$ at the pseudo-critical
temperature we have that $A_{-}{\cal C}(x_m)=A_{\omega}$ and the 
prediction of finite size scaling that 
$\omega=\frac{\alpha}{\nu}$. Taking $L\rightarrow\infty$ at fixed $t$ 
takes the scaling function past the rounding temperature and
scaling gives that for small $x$ that
${\cal C}(x)\sim x^{-\alpha}$ so that we recover 
${\cal C}(T,L)\sim A_{-} t^{-\alpha}$.

In a fully finite system it is not possible to use the expression
\eqref{eq:corr_length_limit} for all $L$ and $T$ and it is also
difficult to apply to numerical data.  Alternative definitions of
correlation lengths, such as the second moment correlation length, are
useful in this contest, see \cite{BarberinDombLebowitz:1983}, but the
scaling analysis is essentially the same.

\subsection{Scaling in terms of $N$}
\label{sec:ScalingN}

Our system has no surface, and all $N_d=3(N^2-1)$ degrees of freedom
are essentially equivalent.  However, in our model the dimensionality,
$d$, of the system is only conjectural (we expect $d=2$ on the fuzzy
sphere side of the transition).  We also do not have access to either
a correlation length or a physical notion of size, $L$.  However, there
is no difficulty in formulating a scaling ansatz in terms of  
the matrix size, $N$.  The essential feature of finite size scaling is
then that the system in the critical regime scales with $N$.

Instead of \eqref{eq:theta_definitionL},
\eqref{eq:lambda_definitionL}, \eqref{eq:omega_definitionL} we use:
\begin{equation}
 |T^{*}(N) - T_c|/T_{c} \sim A_{\overline\theta} N^{-\overline\theta}\,,
  \label{eq:theta_definition}
\end{equation}
for scaling of the rounding temperature with $N$;  
\begin{equation}
  |T_{m}(N) - T_{c}|/T_{c} \sim A_{\overline\lambda}  N^{-\overline{\lambda}}\,,
  \label{eq:lambda_definition}
\end{equation}
for scaling of the shift with $N$ and 
\begin{equation}
   C_{vm}(N) := C_v(T_m(N)) \sim A_{\overline\omega}  N^{\overline{\omega}}.
  \label{eq:omega_definition}
\end{equation} 
for scaling of the peak in the specific heat with $N$.

One can repeat the analysis of the previous section but now using $N$ rather 
than $L$. Scaling suggests that (\ref{ScalingFreeEnergy}) for 
the free energy of the system should be replaced by
\begin{equation}
\lim_{N\rightarrow\infty,T\rightarrow T_c}F(T,N)-N^2f_{\infty}(T_c)
={\cal F}(t N^{\overline\lambda})
\label{ScalingFreeEnergy}
\end{equation}
and we take the scaling variable\footnote{We have set the metric factor here to $1$ for convenience here.} to be $x=t N^{\overline\lambda}$.
Taking the infinite $N$ limit for fixed $t$ gives ${\cal F}(x)\sim \vert x\vert ^{2-\alpha}$ 
and $(2-\alpha)\overline{\lambda}=2$. If we input the theoretical prediction 
for $\alpha=\frac{1}{2}$ we have the further prediction that
\begin{equation}
\overline{\lambda}=\frac{4}{3} .
\label{lambdabar:Prediction}
\end{equation}

Furthermore, once the specific heat has risen
sufficiently above the background values that arise far from $T_c$, we expect
that it has the form
\begin{equation}
C(T,N)=N^{\overline{\omega}}A_{\overline{\omega}}{\cal C}(x)
\label{universalC}
\end{equation}
where ${\cal C}(x)$ is a universal scaling function.
The scaling function ${\cal C}(x)$ must have the behaviour
${\cal C}(x)\sim x^{-\alpha}$ for large $x$. 
Also for $T=T_m(N)$ we require that $x=x_m$ be independent of $N$, 
since $C_{vm}$ is given by (\ref{eq:omega_definition}) and hence 
\begin{equation}
\overline{\omega}=\overline{\lambda}\alpha  =\frac{2}{3}.
  \label{eq:finite_size_scaling_bar}
\end{equation}
The value ${\cal C}(0)$ is a universal number for our system, but it
is difficult to evaluate with any precision due to the difficulties of
accessing this region of parameter space for large matrix sizes.

The relation (\ref{lambdabar:Prediction}) and
\eqref{eq:finite_size_scaling_bar} are derived without reference
to a correlation length or any other notion of distance and 
cannot be used to measure the dimensionality or a characteristic size
of the system. The scaling above is important to us because it
contains only exponents that are directly accessible to our numeric
measurements and can be used to test finite-size scaling in the
context of the current matrix model.

By definition the rounding temperature, $T^*(N)$, is that temperature
where deviations from the asymptotic scaling form begin.  As $N$ is
increased $T^*(N)$ moves closer to the transition temperature.  There
is, however, no unambiguous connection between $T^{*}(N)$ and $N$.
The scaling function ${\cal C}(x)$ should be analytic, aside from its
asymptotic form at large argument, so one can replace $T$ in 
(\ref{universalC})
either with $T^*(N)$ from (\ref{eq:theta_definition}) 
or $T_m(N)$ from (\ref{eq:lambda_definition}) and both
should give a specific heat that diverges with $N$ as 
$N^{\overline{\omega}}$ but with different
amplitudes. Therefore we expect $\theta=\lambda$ and
$\overline{\theta}=\overline{\lambda}$.

If we take the standard relation that $L=aN^{1/d}_d$, which for us gives
$L=N^{2/d}$, assuming the microscopic scale $a=1$ then 
\begin{equation}
  \overline\theta = \theta\frac{2}{d}\,, \quad \overline\lambda = \lambda\frac{2}{d}\,, \quad \overline\omega = \omega\frac{2}{d}\, .
  \label{eq:bar_to_nonbar_exp_relations}
\end{equation}

If we further assume the existence of a single correlation length, $\xi(T)$, 
dominating the critical region then we have
\begin{equation}
  \overline\lambda = \overline\theta = \frac{2}{d\nu} .
  \label{eq:lambda_theta_nu}
\end{equation}
We cannot draw a confident  
conclusion about $\nu$  and $d$ separately as our 
analysis only gives the product $d\nu$
via measurements of $\overline\lambda$ and $\overline\omega$. However,
if we assume that $d=2$, which seems reasonable based on the fuzzy sphere 
as background geometry, we have the prediction for the correlation exponent
\begin{equation}
\nu=\frac{3}{4} .
\end{equation}

We now turn to the numerical measurements.

\section{Near-critical simulation difficulties}
\label{sec:Near-criticalSimulationDifficulties}
In this section we discuss some of the challenges posed in 
a numerical analysis of the properties of the system very close to the
phase transition and describe how they are tackled.

\subsection{Critical Slowing Down}
\label{sec:CriticalSlowingDown}
Critical slowdown is a phenomenon that is typical in numerical
simulations of critical systems.  It is described by the theory of
dynamic critical phenomena (see {\it e.g.}
\cite{Goldenfeld:1992qy,Montvay_Munster} for detailed treatments of
the subject).  When $T\to T_c$ the time needed for a non-equilibrium
system to reach equilibrium grows as does the auto correlation time in
a Monte Carlo simulation of the system.  More detail on the impact of
critical slowdown on our numerics is discussed in the Appendix and
here we merely observe that, for the systems studied in this work, the
critical slowing down of our simulations has significant impact on
systems with $N \ge 100$ and has prevented us from simulating matrices
with $N>110$.\footnote{One can see in Fig.~\ref{pic:omega_fit} that
  the relative error in the measurement of the near-critical specific
  heat grows with the system size. }

\subsection{Excited states}
\label{sec:ExcitedStates}

\begin{figure}
\centering
\includegraphics[scale=0.95]{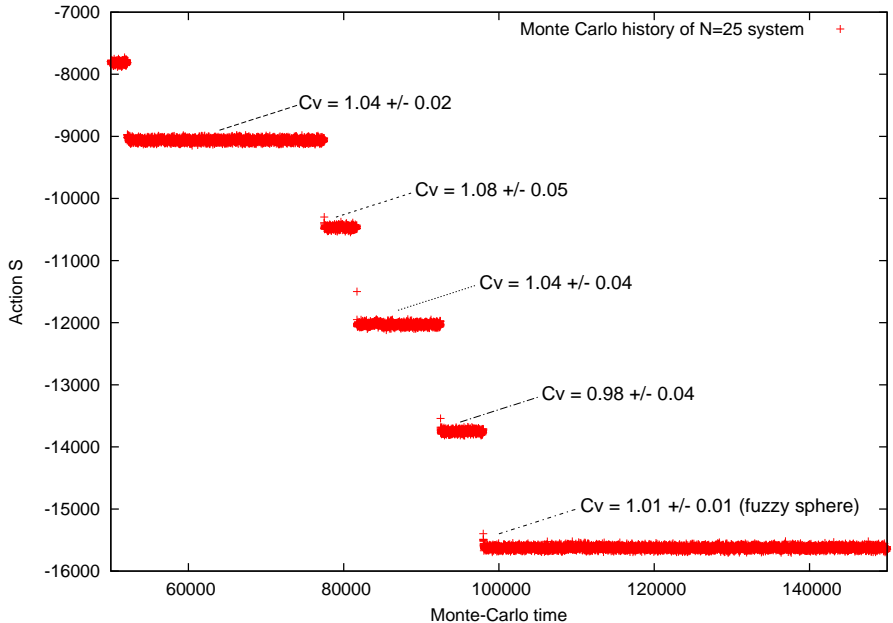}
\caption{Monte Carlo history of an $N=25$ system visiting different excited states, together the corresponding $C_v$ in their specific domains
(for $\gD=5$, i.e.  $T = 0.0016$ and $\frac{T}{T_c}=0.03$).}
\label{pic:excited_states_Cv_extraction}
\end{figure}

Another property to be taken into consideration is the presence of the
excited fuzzy sphere configurations given by
\eqref{eq:excited_states_form}.  As mentioned earlier, those
configurations possess energies which are intermediate between the
commuting matrix phase and the fuzzy sphere phase.  This means that,
in the region where the two phases coexist, we would expect to see
jumps between the ground state and these excited states and between these
excited states and the commuting matrix phase rather than direct transitions
between the commuting matrix phase and the fuzzy sphere ground state.
This expectation is supported by our simulations. Far from the
transition we are able to measure the values of $C_v$ for restricted
ensembles trapped in different excited states and for the fuzzy sphere
ground state. We find that sufficiently far from the transition all
such specific heats are of order $1$ i.e. $C_v\sim 1$ as illustrated
on Fig.~\ref{pic:excited_states_Cv_extraction}.  As the transition is
approached $C_v$ grows and close to the transition distinct states are
no longer observable see Fig.
\ref{pic:transition_point_mc_history_N10}.

\subsection{Energy separation between different phases}
\label{sec:EnergySeparation}

\begin{figure}
\centering
\includegraphics[scale=0.95]{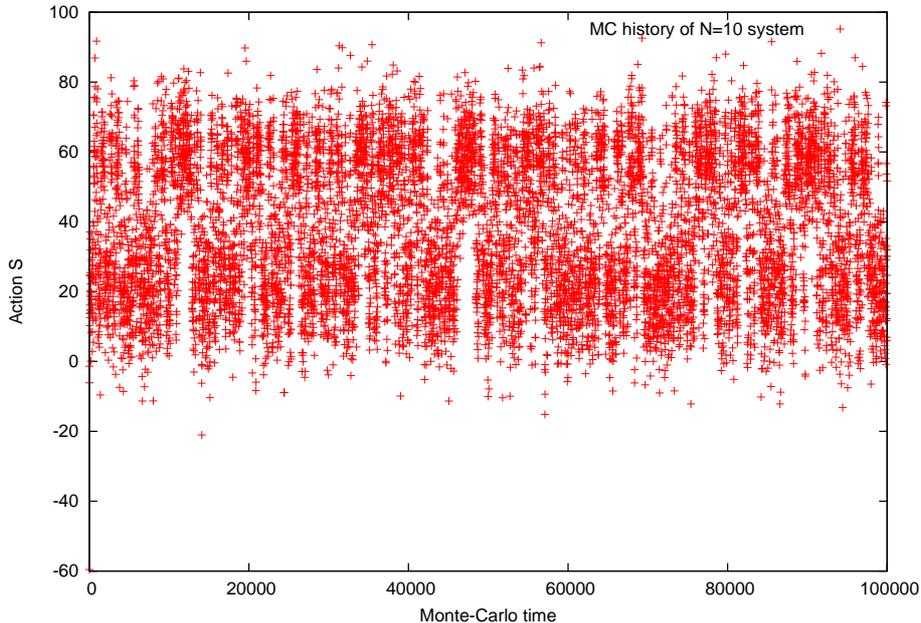}
\caption{Monte Carlo history of an $N=10$ system at $T = 0.0468$, a regime where the two phases coexist. For these values of $N$ and $T$ equation
\eqref{eq:action_exp_value_matrix} 
gives $S_m\approx 75$ while \eqref{eq:semi-classical-S}-\eqref{eq:deltaT} 
give $S_f\approx 20$ and
the action is seen to jump randomly in the vicinity of, and between,  
these two values.}
\label{pic:transition_point_mc_history_N10}
\end{figure}

Ideally in the vicinity of the phase transition the system will jump
between the two phases and if we can get enough Monte Carlo steps, we
will have enough statistics to properly extract the relevant
quantities.  A Monte Carlo history where this occurs is
depicted in Fig.~\ref{pic:transition_point_mc_history_N10}. We can see
that the system spends roughly the same amount of MC time in both
phases, an indication that the system is close to the transition
point.  However for large $N$ the jump between the two phases becomes
rare events, and indeed our numerical studies indicate that this is
already the case for $ N \simeq 12$.  This, in combination with the
asymmetry of the phase transition, makes it very hard to simulate the
system efficiently in the regime where the two phases coexist.

\begin{figure}
\includegraphics[width=1\textwidth]{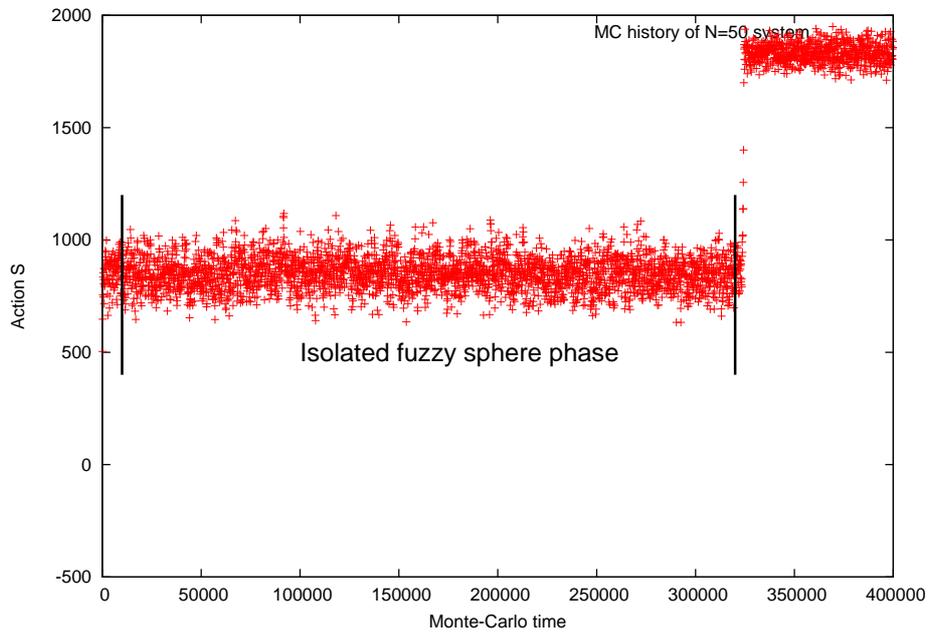}
\caption{Monte Carlo history of an $N=50$ system at $T = 0.0514$
  crossing between the fuzzy sphere phase and the commuting matrix
  phase as an illustration of the restricted ensembles approach.  For
  these values of $N$ and $T$ equation
  \eqref{eq:action_exp_value_matrix} gives $S_m\approx 1875$ while
  \eqref{eq:semi-classical-S}-\eqref{eq:deltaT} give $S_F\approx 850$.
  The vertical bars show the fuzzy sphere domain, if we restrict our
  measurements only to this region we can extract observables in the
  fuzzy sphere phase. These are what we call restricted ensemble
  measurements.}
\label{pic:Isolated_phase}
\end{figure}

One tactic to handle this problem is to perform a \emph{cold start} on
the Monte Carlo runs, so that the phase transition is always approached 
from the low temperature side.  This biases the system toward the fuzzy
sphere phase, but has the advantage of giving reproducible results.
An example is shown in Fig.~\ref{pic:Isolated_phase}, with $N=50$ at $T=0.0514<T_c$. The system is below the critical temperature but, while the value of the
action is compatible with a fuzzy sphere configuration for quite some
Monte Carlo time, it suddenly jumps to a commuting matrix configuration. 
Once in the matrix phase configuration, it remains there as fluctuations
are too small to get it back.

\subsection{Comments on the algorithm}
When starting the system in a zero field configuration, and using a
simple Monte Carlo simulation, the system tends to get stuck in the
zero action local minimum. The typical configuration that such simulations
produce is one where two of the matrices have zero
eigenvalues while those of the third matrix acquires non-zero
eigenvalues which are distinct from one another. The matrix that is
started first in the simulation is the one whose eigenvalue become
non-zero and non-degenerate.  A hybrid Monte Carlo simulation is
necessary to overcome this difficulty and for our study we use the
hybrid Monte Carlo algorithm of \cite{Duane:1987de}.

It is also tempting to perform simulations for the system by first
diagonalising one of the matrices. This results in a Vandermonde Jacobian 
from the change of variables whose logarithm is included in the effective 
action. This algorithm is quite efficient for some simulations but simulations 
with a hot start find it much more difficult to relax to a fuzzy 
sphere phase even for very low temperatures and relatively large matrices. 
This is because separation of the eigenvalues from one another 
must filter from the outer eigenvalues inwards which tends to be very slow.
To avoid any such difficulties we have chosen to use the direct approach 
and our simulations are performed with a hybrid Monte Carlo algorithm 
in which all matrices are treated on an equal footing.

\section{Numerical measurement of critical exponents and finite size scaling}
\label{sec:NumericalMeasurement}

A direct analysis of the specific heat data, in the immediate vicinity
of the critical point, would involve a four parameter fit
\begin{equation}
C_v(T)=C_{0-}+A_{-}(T_c-T)^{-\alpha}
\end{equation}
where the data are fit to obtain $C_{0-}$, $A_{-}$, $T_c$ and $\alpha$.
This, however, involves large errors as there are so many parameters
and the specific heat data for $T$ very close to $T_c$ involves finite
size effects which mean that the finite size scaling function 
${\cal C}(x)$ enters the picture. The first step is therefore a the
determination of the critical temperature.

\subsection{Estimating $T_c$ and $\overline{\lambda}$ and $\overline{\omega}$}
\label{sec:Tc}
\begin{figure}
\includegraphics[scale=0.95]{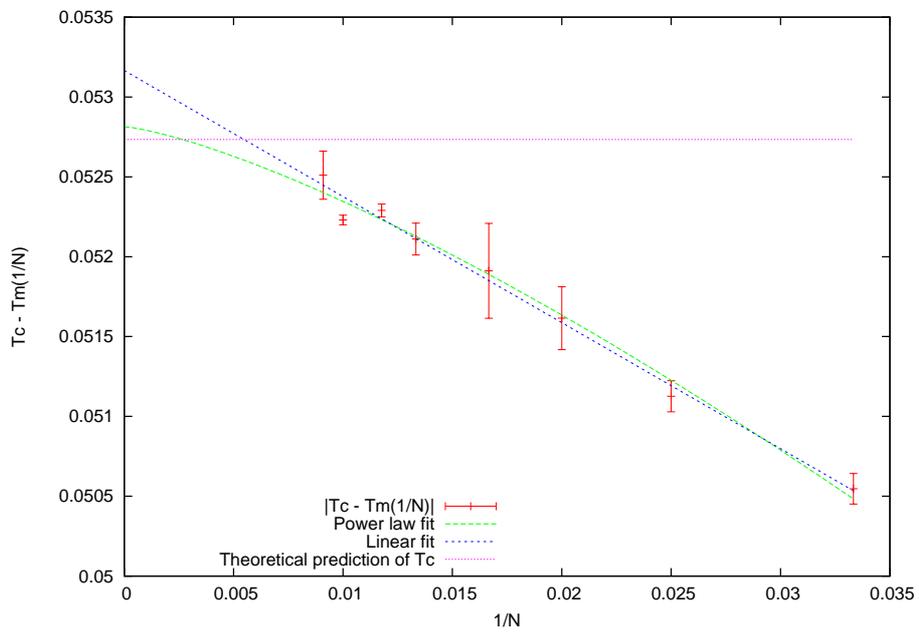}
\caption{Extrapolation of $T_m(N)$ for $\frac{1}{N}\to 0$. 
The value at $0$ corresponds to $T_m(\infty)$ which is $T_c$. The theoretical prediction is given by the blue dashed line. The green line is a three parameter fit to equation (\ref{eq:lambda_definition}) with $T_c=0.0531\pm 0.0003$, 
$A_{\overline\lambda}=1.8\pm 1.5$, $\overline\lambda = 1.1\pm 0.3$}
\label{pic:critical_point_extrapolation}
\end{figure}

A precise determination of the critical temperature is necessary for
the evaluation of the specific heat exponent $\alpha$ and the shift
exponent $\overline\lambda$. 
For a given finite $N$ our best estimate of the critical
temperature is the pseudo-critical temperature $T_m(N)$.
We therefore first analyse $T_m(N)$ and endeavour to extract $T_c$ 
from the limit of $T_m(N)$ for $N$ going to infinity by
fitting it as a function of $N$ to the shift scaling form
(\ref{eq:lambda_definition})
\begin{equation}
 T_{m}(N) =T_c(1-A_{\overline\lambda}  N^{-\overline{\lambda}})\, .
\end{equation}
In Fig.~\ref{pic:critical_point_extrapolation} we present simulation 
our data for $T_m(N)$.

Visual inspection of the data suggests a linear fit,
i.e. $\overline{\lambda}=1$.  Linear regression on the data in 
Fig. \ref{pic:critical_point_extrapolation} gives $T_c= 0.0532\pm0.0001$ and
$A_{\overline{\lambda}}=1.5\pm0.1$ but assumes
$\overline{\lambda}=1.0$.  However, the scaling ansatz suggests that
we should look for a three parameter fit. When we perform such a three
parameter fit we get $T_c= 0.0531\pm 0.0003$,
$A_{\overline{\lambda}}=1.8\pm 1.5$ and $\overline{\lambda}=1.1\pm0.3$ and 
the resultant critical temperature from both fits is largely unchanged. 

For the three parameter fit the error in the amplitude $A_{\overline\lambda}$ is 
rather large so it is desirable to fix some of the parameters. Since our principal goal is to check scaling we need to measure $\overline{\lambda}$ rather 
than assume it. The measurements of $T_c$ from both the three parameter fit
and the linear one broadly agree with the theoretical prediction 
$T_c={(\frac{3}{8})}^3\simeq 0.0527344$, suggesting
that we constrain the fit so that $T_c$ is fixed to be this number.
With $T_c$ so constrained we then find a two parameter gnuplot fit
gives $A_{\overline{\lambda}}= 5.2\pm 1.6$ and
$\overline\lambda=1.41\pm 0.08$. 

Finally since we have a prediction for the exponent
$\overline{\lambda}$, see (\ref{lambdabar:Prediction}), fixing both $T_c$ and
$\overline{\lambda}=\frac{4}{3}=1.333$ gives $A_{\overline{\lambda}}=
3.9\pm0.1 $ suggesting perhaps that $A_{\overline{\lambda}}=4$.  We
conclude that the data are quite consistent with the theoretical
estimate though a linear fit is also consistent with our measurements.

\begin{figure}
\includegraphics[scale=0.95]{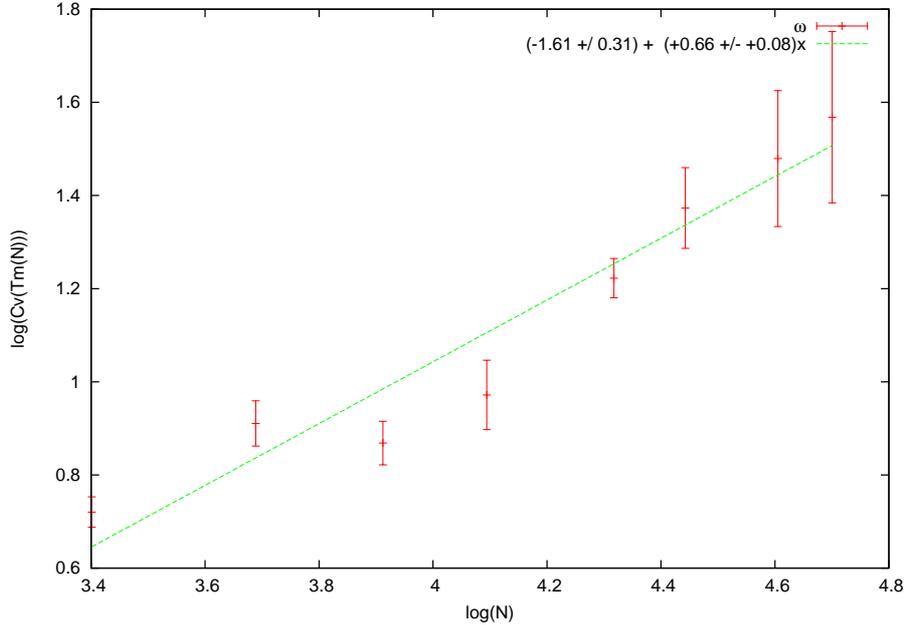}
\caption{A plot of $C_v(T_m(N))$ as a function of $N$, on a double-logarithmic scale, and least squares fit to data. The slope of the linear fit corresponds to $\overline\omega$ in \eqref{eq:omega_definition} for $C_v$ and gives $\bar\omega=0.66\pm0.08$.}
\label{pic:omega_fit}
\end{figure}

We can similarly analyse the maximum of the specific heat $C_{vm}(N)$.
A linear fit assumes $\overline{\omega}=1$ and is best interpreted as
an estimation of $C_{0-}$, our data from such a linear fit gives 
$C_{vm}=0.8\pm0.2+(0.035\pm0.002)N$  which gives a value of 
$C_{0-}$ consistent with the theoretical prediction of 
$\frac{29}{36}\simeq0.8055$.  A three parameter fit gives 
$C_{vm}=(1.7\pm0.3)+(0.001\pm.002)N^{1.7\pm0.4}$,  but the amplitude is 
very small and has large errors. Our scaling ansatz suggests that we 
should look for 
a fit $C_{vm}(N)=A_{\overline\omega}N^{\overline\omega}$ which is best extracted from 
a log-log plot. Such a log-log plot is shown in Fig. \ref{pic:omega_fit}
and gives $A_{\overline\omega}=0.21\pm0.06$ and $\overline{\omega}=0.66\pm0.08$ 
which is surprisingly close to the theoretical prediction 
$\overline{\omega}=\frac{2}{3}$.

Summarising our data: When the critical temperature is assumed to be the 
theoretical value we find with two parameter fits the 
amplitudes and exponents in \eqref{eq:lambda_definition} 
and \eqref{eq:omega_definition} are given by
\begin{equation*}
  \frac{T_{c} - T_{m}(N)}{T_c} \sim (5.2 \pm 1.6)  N^{-1.41 \pm 0.08} 
\end{equation*}
\begin{equation*}
  C_v(T_m(N))\sim (0.21 \pm 0.06)  N^{0.66 \pm 0.08}. 
\end{equation*}
where we have chosen to prefer the direct fit for 
$T_m(N)$ and the logarithmic fit for $C_{vm}(N)$.
The basic data used are shown in
Fig.~\ref{pic:critical_point_extrapolation} and \ref{pic:omega_fit}

The data yield the exponents $\overline\lambda = 1.41 \pm 0.08$ and
 $\overline\omega = 0.66 \pm 0.08$ and the scaling relation $\alpha={\overline{\omega}}/{\overline{\lambda}}$, see eqn. (\ref{eq:finite_size_scaling_bar}), 
predict $\alpha=0.47\pm0.06$  which within errors is consistent 
with $\alpha=0.5$.

\subsection{Direct measurement of $\alpha$ from collapsed data}
\label{sec:DirectMeasurement}

As argued in the previous section the numerical estimate for $C_{0-}$
in \eqref{eq:CvT} when extracted from a linear fit to the specific 
heat maximum, is in good agreement with the theoretical prediction
\eqref{eq:Cv_critical_behavior} $C_{0-}=\frac{29}{36}$. It is then reasonable 
to take this as the input value of $C_{0-}$ and endeavour to extract 
the amplitude and exponent $\alpha$ from the data. 

However, data for a given fixed $N$  are not
very satisfactory for the estimation of $A_{-}$ and $\alpha$ 
since any such data have important finite size
corrections for finite $N$, i.e.  the presence of values of the
specific heat taken at temperatures $T < T^{*}(N)$ when extracting the
exponent from single matrix size data.  As we pointed out earlier, the
point $T = T^{*}(N)$ is hard to detect. Since the rounding temperature
depends on $N$, we can do better by combining different values of
$N$. Deviations due to the rounding effects are significantly
diminished by averaging over the specific heat for different
$N$. 

\begin{figure}
\centering
\includegraphics[scale=0.95]{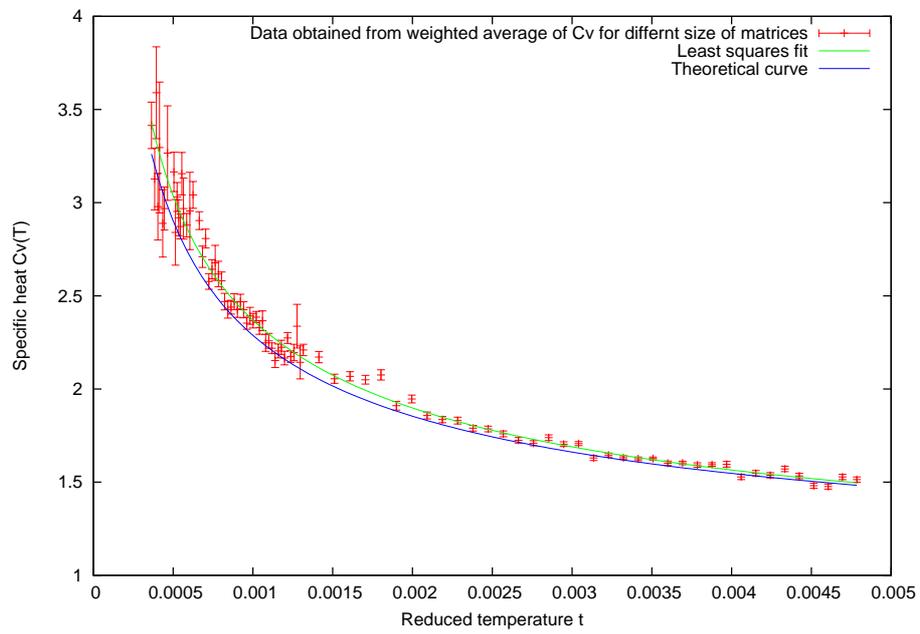}
\caption{The figure shows the weighted averaged $C_v$ obtained from 
different values of $N$ ranging from 30 to 110. A three parameter fit to the 
data with $C_v=C_{0-}+A_{-}(T_c-T)^{\alpha}$ gives $C_{0-}=0.76\pm 0.09$, 
$A_{-}= 0.051\pm 0.017$ and $\alpha=-0.50\pm 0.04$.}
\label{pic:Cv_collapsed}
\end{figure}

Scaling suggests that, for a fixed value of the temperature lower than
$T^{*}(N)$, the values of the specific heat should be consistent for
different $N$, provided $N$ is large enough, and our measurements
verify this expectation. So we treat such values as independent
measurements of $C_v(T)$ and then take the weighted average over such
values.  If we have a number of independent measurements of $C_v$,
labeled by a discrete index $i$ for different values of matrix size
$N_i$, then the weighted average, $C_v(T)$, is defined to be
\begin{equation}\label{eq:WeightedCv}
  C_v(T) = \frac{\sum_i \frac{1}{\sigma_{N_i}^2} C_v(T, N_{i})}{\sum_i \frac{1}{\sigma_{N_i}^2}}
\end{equation}
where $\sigma_i$ is the uncertainty in measurement $i$.
Fig.~\ref{pic:Cv_collapsed} plots a weighted average $C_v$ using data 
corresponding to values of $N$ ranging from 30 to 110.

Fig. \ref{pic:Cv_collapsed} shows combined data for different $N$. 
When we approximate the near-critical behavior of the specific as 
$C_v \sim C_{0-} + A_{-}|T_c - T|^{-\alpha}$ 
and use a three parameter fit we find 
$C_{0-}= 0.76\pm 0.09$, $A_{-}= 0.051\pm 0.017$ and $\alpha=-0.50\pm
0.04$. All three parameters are in very good agreement with
(\ref{eq:Cv_critical_behavior}).  One might expect $C_{0-}$ to be the least
sensitive of the parameters in the approach to the singularity, and it 
is tempting to set $C_{0-}$ to the theoretical background value and
refine the estimate of $A_{-}$ and $\alpha$.  When this is done we
find $A_{-}= 0.043\pm 0.0014$ and $\alpha=0.5197\pm 0.0054$. If we also 
set  $\alpha=\frac{1}{2}$ for a one parameter fit 
we get $A_{-}=0.0487\pm 0.0002$ which suggests that the true 
value is indeed $A_{-}=\frac{3}{64}=0.046875$.
It should be noted that the form (\ref{eq:Cv_critical_behavior}) is 
only asymptotic and has additional corrections. Also our data still has 
finite size effects included.   As a final estimate 
we set $C_{0-}=\frac{29}{36}$ 
and perform a two parameter direct fit to the data  
we get $A_{0-}=0.047\pm0.001$ and $\alpha=0.51\pm 0.01$.
In summary we believe that our data gives reasonable evidence
that (\ref{eq:Cv_critical_behavior}) captures the true large $N$ 
behaviour of the system.

On the high temperature side of the transition we find no divergence
of the specific heat, and our numerical measurements show good
agreement with the value $C_v = \frac{3}{4} = const$.  To conform to
the standard scenario of critical phenomena there are two
alternatives: either $A_{+} = 0$ or $\alpha = 0$ for $T > T_c$.  If we
assume the critical exponent is equal on the two sides of the
transition we are led to the conclusion that $A_{+} = 0$. Then we have
the universal ratio $U_{0} = A_{+} / A_{-} = 0$, the amplitudes
$A_{+}$ and $A_{-}$ are system and/or interaction dependent, but $U_0$
is universal.  This provides important information for determining the
universality class of the system.

\section{Conclusions and outlook}
\label{sec:Conclusions}

One of the key features of our simulations is that the matrix phase
configurations tend to be extremely stable.  Fluctuations around these
are small, they have a restricted ensemble specific heat of
$C_v=\frac{3}{4}$ in comparison with the fuzzy sphere background whose
minimum specific heat is $C_v=1$. This means that the decay of matrix
phase configurations becomes very unlikely even very close to the
transition. The small fluctuation analysis of section
\ref{Sec:CommutingMatrixPhase} suggests that the principal mode of
decay of such configurations is via the negative eigenvalue
identified in section \S \ref{Sec:CommutingMatrixPhase} and due to the  
Myers term. The largest
eigenvalue separation of matrix phase configurations is
$\Delta_{ij}^a=2R$ and such configurations have eigenvalues
described by the parabolic distribution (\ref{parabolicdistribution})
so one would expect the matrix phase to become unstable if on average
the eigenvalues of one of the matrices are too close together. Taking
the expectation value of (\ref{eq:instability_condition}), assuming a 
parabolic distribution of the eigenvalues, one 
obtains that the matrix phase becomes unstable for
$g^2>g_m^2\simeq\frac{1}{3}\frac{<\Delta_{ij}^2>}{4}=\frac{R^2}{10}$, where 
the factor of $1/3$ comes from averaging over the number of matrices. 
For $\tilde{g}$
independent of $N$ this gives
$\tilde{g}_m\simeq\frac{R\sqrt{N}}{\sqrt{10}}$. So the matrix phase becomes
more and more stable as $N$ increases.  Noting that numerically
$R=2.0$ we find that $\tilde{g}_m\simeq0.632 \sqrt{N}$ so that for $N<11$ we
have $\tilde{g}_{m}(N) < {(\frac{8}{3})}^{\frac{3}{4}}$ while for
$N\ge11$ it is larger.  For large $N$ and fixed $\tilde{g}$ we
therefore expect simulations to get trapped in the matrix phase if the
simulation ever falls into such configurations and we further expect
that for matrix sizes much larger that $N=11$ it will be virtually
impossible to escape from the matrix phase.  This is precisely what we
observe in simulations.  We find that with effort one can escape from
the matrix phase for $N\leq14$ but our simulations have great difficulty escaping for $N=15$ and do not escape for
larger $N$.  In fact older numerical simulations \cite{Azuma:2004zq}
give the instability of the matrix phase as $g_{m}=0.66$ which is
consistent with our observations.

Simulations on small matrix sizes then have quite a different
character to those for large matrix sizes. They exhibit fluctuations
that make transitions between the two types of typical
configuration--the fuzzy sphere and the matrix phase. For larger $N$
such fluctuations are absent and one is forced to take data in a
restricted ensemble where the fluctuations are either around the fuzzy
sphere or around the matrix phase. 

In this work we have not endeavoured to study small enough systems 
where fluctuations between the matrix phase and the fuzzy sphere 
are possible. We have rather concentrated on larger matrix sizes
as our goal is to check finite size scaling of fluctuations with $N$. 
Our simulations therefore probe the scaling properties of fluctuations in the
fuzzy sphere phase. The fluctuations grow with $N$ and diverge in the
large $N$ limit at a critical temperature with the large $N$ fuzzy
sphere becoming unstable at $\tilde{g}\simeq{(\frac{8}{3})}^{3/4}$ corresponding
to a critical temperature $T_c\simeq {(\frac{3}{8})}^{3}$. 
These fluctuations, since they do not probe the matrix phase configurations, 
are what we call restricted ensemble fluctuations. 

It may be that the true thermodynamically stable ground state of the system 
is in fact the matrix phase, and simulations on small
matrix sizes tend to suggest this, however, in the large $N$ limit and
for $T<(\frac{3}{8})^3$ fluctuations never escape from the vicinity of
the fuzzy sphere. If the fuzzy sphere phase is only a local minimum and
not a global minimum in the vicinity of this transition,  
then it would be clear why the 
transition appears to be one-sided---the interpretation would be that the 
system is trapped in a local minimum which is stable due to the large $N$ 
limit and the observed transition would be due to the configurations 
eventually escaping over the barrier once the temperature is high enough. 
The transition would then be very similar 
to the two dimensional quantum gravity transition discussed in the 
matrix model literature \cite{DiFrancesco:2004qj} which  also has 
a one sided transition. The quantum gravity system is exactly solvable
and in the case of four valent planar graphs where the potential is 
\begin{equation}
V(\Phi)=N Tr(\frac{1}{2}\Phi^2-\frac{g}{4}\Phi^4)
\end{equation}
the eigenvalue distribution is given by
\begin{equation}
\rho(x)=\frac{1}{2\pi}(1-g\frac{a^2}{2}-g x^2)\sqrt{a^2-x^2}\quad{\rm with}\quad a^2=\frac{2}{3g}(1-\sqrt{1-12g})
\end{equation}
and the specific heat by 
\begin{equation}
C_v=\frac{1+54g^2-(1+6g)\sqrt{1-12g}}{216 g^2} .
\end{equation}
By rescaling  $\Phi\rightarrow \varphi/\sqrt{g}$ one can rewrite 
$V(\Phi)=\frac{N}{g}Tr(\frac{1}{2}\varphi^2-\frac{1}{4}\varphi^4)$ 
and we can be interpreted the coupling as temperature, $g=T$. 
The system has a critical temperature $T_c=g_c=\frac{1}{12}$ and a
non-analytic specific heat $C_v=\frac{11}{12}+\sqrt{12}\sqrt{T_c-T}+\cdots$
corresponding to specific heat exponent $\alpha=-\frac{1}{2}$. As in
our case, a restricted ensemble, where the system is confined to the
well near the origin, will capture this behaviour. The critical point is
when the eigenvalues spread enough to spill over the barrier.

As far as our simulations are concerned we start them in, or
near, the fuzzy sphere ground state and, for $T\leq {(\frac{3}{8})}^3$
and sufficiently large $N$, they almost never escape from this. 
Our principal observations are restricted to this regime and in this
context we have demonstrated that, despite the non-locality of
matrix actions and the absence of a characteristic size, finite-size
scaling, with matrix size $N$, is still valid.

We have measured the critical temperature and the specific heat
critical exponent, $\alpha$, along with the finite-size scaling
exponents $\overline\omega$ and $\overline\lambda$ near the phase
transition.  Our numerical analysis is compatible with the hypothesis
that finite-size scaling is valid, see \S \ref{sec:ScalingN} for details.

The values obtained for these exponents are new and we know of no other system
with these exponents which suggest that the model under study belongs
to a new universality class. We expect that the critical exponents are
universal within the class of models with fuzzy sphere geometries
evaporating.

Since we have not measured a correlation length and associated critical exponent we cannot establish that our shift exponent $\overline\lambda$ is related to 
the the correlation exponent. The measurement of a correlation length exponent 
is a non-trivial exercise in this context. 

In order to verify the relations \eqref{eq:lambda_theta_nu},
i.e. $\overline{\lambda}=\frac{2}{d\nu}$, one needs to measure the
exponent $\nu$ for some appropriately defined effective correlation
length $\xi(T)$.  It seems plausible that a correlation length could
be extracted by studying the fall-off of two point correlators, but
this is left to future work.

{\bf Acknowledgements:} We wish to thank Thomas Kaltenbrunner for many 
helpful discussions.

\appendix
\section{Critical Slowing Down}
\label{sec:Appendix}

In numerical calculations critical slowdown manifests itself as a 
growth in the number of consecutive measurements of an observable $P$
which are correlated as the transition is approached, i.e. the 
autocorrelation time $\tau_P$ grows as the transition is approached.

Consider a simulation consisting of $N_{mc}$ Monte Carlo steps
producing a set of measurements $\{P_1,\ldots P_{N_{mc}} \}$. If $\tau_P \le 1$
an expectation value $<P>$ can be computed and assigned 
an uncertainty $\sigma \sim 1/\sqrt{N_{mc}}$.
When $\tau_P > 1$ this error estimation is too optimistic,
because the measurements are not fully independent, and
a better estimation is given by $\sigma_{\tau_P} \sim 1/\sqrt{N_{mc}/(2\tau_P)}$.
The correlated data set is effectively equivalent to an uncorrelated data set consisting of $N_{mc}/(2\tau_P)$ measurements. 

The autocorrelation time associated with an observable $P$
is expected to be  governed by the correlation length
and near a critical point should behave as
\begin{equation}
  \tau_P \sim \xi(T)^{d + z(P)},
  \label{eq:dynamic_scaling}
\end{equation}
with dynamical scaling exponent $z(P)$, which
is algorithm and observable dependent.
One aim in designing an efficient algorithm is to reduce $z(P)$.

For a critical matrix model with size $N$ we would expect, assuming
\eqref{eq:corr_length_exponent}, \eqref{eq:theta_definition}, 
\eqref{eq:lambda_definition} and
\eqref{eq:lambda_theta_nu},
\begin{equation}
  \tau_P \sim N^{2 + z(P)\frac{2}{d}}\,.
  \label{eq:matrix_dynamic_scaling}
\end{equation}

 In principle the autocorrelation time for an infinite data set is computed using the series 
\begin{equation}
  \tau_P = \frac{1}{2} \sum_{n = -\infty}^{\infty} \sum_{\tau = -\infty}^{\infty} \frac{<(P_n - <P>)(P_{n+\tau} - <P>)>}{<(P_{n} - <P>)^2>}\,,
  \label{eq:autocorrelation_time_series}
\end{equation}
but in practice the sum $\sum_{\tau = -\infty}^{\infty}$ must obviously be truncated to $\sum_{\tau = -\tau_0}^{\tau_0}$
with $\tau_0$ finite and in general the obtained value for 
$\tau_P$ depends on $\tau_0$.
Clearly the truncated version of \eqref{eq:autocorrelation_time_series} can only be sensitive to autocorrelation times $\lesssim \tau_0$
so,  if $\tau_0 \ll N_{mc}$, $\tau_p$ might be underestimated for systems with severe autocorrelation. On the other hand, if $\tau_0 \sim N_{mc}$,
the convergence of \eqref{eq:autocorrelation_time_series} becomes very poor. 
We can not rely only on the above expression to determine $\tau_P$.

In our analysis we allow for autocorrelations using the \textit{jackknife}
procedure, see {\it e.g.} \cite{Montvay_Munster}, 
which computes
the uncertainty taking into account the autocorrelation of the data. 
As a consistency check we compute the autocorrelation using the expression \eqref{eq:autocorrelation_time_series}, 
with empirically chosen $\tau_0$ and then we compare the result with $\tau_P$ as determined by the 
jackknife procedure.

\vfill\eject

\end{document}